\documentclass{article}

\usepackage{PRIMEarxiv}
\usepackage[utf8]{inputenc} 
\usepackage[T1]{fontenc}    
\usepackage{hyperref}       
\usepackage{url}            
\usepackage{booktabs}       
\usepackage{amsfonts}       
\usepackage{nicefrac}       
\usepackage{microtype}      
\usepackage{lipsum}
\usepackage{fancyhdr}       
\usepackage{graphicx,comment}       
\graphicspath{{media/}}     

\title{TOSHFA: A Mobile VR–Based System for Pose-Guided Exercise Rehabilitation for Low Back Pain
\thanks{This study has been approved by the IRB Approval Committee at VRapeutic Inc. and the Research Ethics Approval Committee of the Electronics and Communications Engineering Department at the Arab Academy for Science, Technology, and Maritime Transport (VRAP.MO/07).}
}

\author{
  Amin Mohamed\\
  CSAI School, Zewail City of Science and Technology  \\
  \texttt{s-amin.mohamed@zewailcity.edu.eg} \\
  \And
  Hamza Abdelmoreed\\
  CSAI School, Zewail City of Science and Technology  \\
  \texttt{s-hamza.fekry@zewailcity.edu.eg} \\
  \And
  Mohamed Ehab \\
  CSAI School, Zewail City of Science and Technology  \\
  \texttt{s-mohamed.ehab@zewailcity.edu.eg} \\
  \And
  Youssef Shawky \\
  CSAI School, Zewail City of Science and Technology  \\
  \texttt{s-s-yousef.shawky@zewailcity.edu.eg} \\
  \And
  Mayada Hadhoud \\
  CSAI School, Zewail City of Science and Technology \\
  Faculty of Engineering , Cairo Univeristy\\ 
\texttt{mhadhoud@zewailcity.edu.eg} \\
  \And
  Ahmad Al-Kabbany \\
  VRapeutic Inc. \\
  Multimedia Interaction and Communication Lab \\
  Wearables, Biosensing, and Biosignal Processing Research lab \\
  Arab Academy for Science and Technology \\
  \texttt{alkabbany@ieee.org, alkabbany@aast.edu} \\
}

\begin{document}
\maketitle

\begin{abstract}
Low back pain (LBP) is a pervasive global health challenge, affecting approximately 80\% of adults and frequently progressing into chronic or recurrent episodes. While exercise therapy is a primary clinical intervention, traditional at-home programs suffer from low adherence rates and the absence of professional supervision. This study introduces TOSHFA, an accessible mobile VR-based rehabilitation system that bridges this gap by combining computer vision with affordable hardware. The system utilizes a laptop webcam to perform real-time pose estimation via the MediaPipe framework, tracking 33 skeletal landmarks to provide immediate biofeedback. This data is streamed via low-latency UDP protocols to a smartphone mounted in a cardboard-style VR headset, where patients interact with a gamified 3D environment.
A pilot study with 20 participants evaluated the system's performance and user engagement. Quantitative results yielded a mean System Usability Scale (SUS) score of 47.4, indicating marginal usability and a need for interface optimization. However, Game Experience Questionnaire (GEQ) data revealed high scores in positive affect and enjoyment, suggesting that the gamification elements—such as coin rewards and streak tracking—successfully maintained user motivation despite technical friction. These findings validate the feasibility of a smartphone-based telerehabilitation model and establish a technical foundation for future clinical trials involving multi-exercise protocols.
\end{abstract}


\section{Introduction}

Lower back pain (LBP) is a prevalent global health challenge, affecting approximately 80\% of adults at least once in their lives. Approximately 20\% to 30\% of this population develops chronic LBP, leading to significant disability and escalating healthcare costs. Furthermore, 60\% to 80\% of patients experience recurrent episodes within a single year. Clinical guidelines, such as those from the Mayo Clinic~\cite{mayoclinic_backpain_prevention}, suggest that targeted exercises—including seated lower back rotation stretches—can significantly improve flexibility, reduce stiffness, and enhance spinal mobility. However, the efficacy of these interventions is heavily dependent on patient adherence. Moreover, performing these movements inaccurately significantly increases the risk of exacerbating the condition, making real-time, constant feedback a vital component of successful rehabilitation outcomes.  

While LBP is traditionally managed through physical therapy, several systemic barriers limit the effectiveness of conventional treatments. A limited supply of therapists often results in protracted wait times for clinical appointments, which is further compounded by the high cost of frequent in-person sessions. Consequently, many patients transition to at-home programs; however, these often fail due to a lack of consistent follow-up, difficulty in comprehending exercise instructions, or diminished motivation. Additionally, the geographical and temporal separation between physicians and patients post-consultation prevents ongoing monitoring and corrective feedback. These challenges underscore the urgent need for accessible, cost-effective solutions that provide the regular feedback necessary to maintain patient engagement.  

Recent advancements in Virtual Reality (VR) and computer vision offer a transformative approach to rehabilitation by enabling high levels of interaction and real-time biofeedback. Markerless pose estimation frameworks, such as MediaPipe, have demonstrated the ability to accurately detect movement patterns, establishing them as a viable foundation for home-based telerehabilitation. While VR-based systems have been shown to enhance patient motivation by simulating the engaging experience of a clinic, adoption is often hindered by the high cost of dedicated hardware, such as the Oculus Quest. To address this lack of accessibility, researchers are increasingly looking toward smartphone-based VR and cardboard headset systems, which offer affordable and scalable alternatives to high-end equipment.

Despite the recognized benefits of Virtual Reality (VR) in clinical settings, many current solutions fail to align with evidence-based LBP guidelines regarding exercise design. Numerous systems utilize VR primarily to enhance user enjoyment, often neglecting the critical components of an effective LBP rehabilitation program, such as precise body positioning, anatomically appropriate ranges of motion, and specific exercise durations. Furthermore, the majority of existing VR rehabilitation games overlook sophisticated game-based motivation techniques, relying instead on traditional usability metrics like the System Usability Scale (SUS). In contrast, this study integrates real-time biofeedback with intentional gamification to foster superior user engagement and sustained motivation. Moreover, the dependence of most current VR options on expensive, dedicated headsets remains a significant barrier to accessibility for patients seeking home-based care.  

To address these gaps, this paper presents a clinically grounded, pose-guided exercise system specifically engineered for LBP. The system adheres to established clinical practice guidelines to ensure the safe and accurate execution of prescribed rehabilitation movements. By leveraging a low-cost mobile VR architecture—integrating a standard laptop webcam with a smartphone and a cardboard-style headset—we provide a solution that is simultaneously accessible, affordable, and suitable for independent at-home use. The contributions of this article can be summarized as follows:

\begin{enumerate} 

\item Development of a Clinically Grounded Framework: The design of a pose-guided exercise system for LBP that integrates established clinical guidelines to ensure patients can accurately and successfully complete prescribed rehabilitation movements. 

\item Low-Cost Telerehabilitation Architecture: An affordable, home-based rehabilitation solution that leverages mobile VR technology to bypass the need for expensive, dedicated head-mounted displays. 

\item Comprehensive User Experience Evaluation: A pilot study involving 20 participants that utilizes a dual-metric approach—combining the System Usability Scale (SUS) for pragmatic usability and a customized Game Experience Questionnaire (GEQ) for hedonic engagement. 

\item Future Clinical and Technical Roadmap: A strategic research framework for expanding the system to support multi-exercise and multi-session protocols, providing a foundation for future large-scale clinical trials. \end{enumerate}

The rest of this paper is organized as follows: Section 2 reviews related work in VR-based rehabilitation for LBP, Section 3 provides an overview of the system architecture, Section 4 describes the methods used in the study, Section 5 presents the results and discusses the findings, and Section 6 concludes the paper with suggested future work.

\section{Related Work}

\subsection{VR-Based Rehabilitation for Low Back Pain and Musculoskeletal Disorders}
Technology-assisted rehabilitation has been increasingly explored to improve access to care and enhance adherence to therapeutic exercise, particularly for chronic low back pain (LBP) and musculoskeletal disorders. Virtual reality (VR)–based rehabilitation has received growing attention due to its potential to provide immersive, interactive environments that encourage active participation and repeated practice.

Systematic evidence supports the clinical potential of VR interventions for LBP. Brea-Gómez et al. \cite{brea2021} conducted a systematic review and meta-analysis of randomized clinical trials and reported that VR-based rehabilitation significantly reduced pain intensity and kinesiophobia while improving functional outcomes in adults with chronic LBP. Similarly, Li et al. \cite{li2024} summarized evidence from randomized controlled studies and found that VR-based training produced meaningful improvements in pain and disability compared with conventional rehabilitation approaches.

Beyond LBP, recent reviews indicate that VR interventions may also benefit a broader range of musculoskeletal pain conditions. Opara Zupančič and Šarabon \cite{opara2025} reviewed the state of VR applications for musculoskeletal conditions and associated chronic pain, noting positive effects on pain reduction and physical performance across multiple body regions, including the spine. However, the authors emphasized substantial heterogeneity in intervention design, hardware platforms, and outcome measures.

Research has also explored the integration of VR within remote and home-based rehabilitation frameworks. Garofano et al. \cite{garofano2025} reported that VR-enhanced remote rehabilitation programs incorporating sensor-based feedback improved pain outcomes, functional mobility, and adherence in individuals with chronic LBP. Collectively, these studies suggest that while VR-based rehabilitation is promising, scalable and accessible system designs remain critical for effective unsupervised or remote use.

\subsection{Deep Learning for Pose Estimation and Movement Analysis in Rehabilitation}
Markerless pose estimation provides a pathway to quantify rehabilitation exercise technique without body-worn sensors. Sarafianos et al. \cite{SARAFIANOS20161} reviewed 3D pose estimation methods and identified challenges that remain salient in home settings, including viewpoint variation, limited depth cues, and self-occlusion. Real-time pipelines have since matured for consumer hardware: Lugaresi et al. \cite{lugaresi2019mediapipe} introduced MediaPipe as an efficient perception framework, Bazarevsky et al. \cite{bazarevsky2020blazepose} proposed BlazePose for on-device real-time body pose tracking, and Cao et al. \cite{8765346} described OpenPose as a widely used baseline for real-time multi-person 2D pose estimation.

Recent practical comparisons have focused on the accuracy–latency trade-off that determines whether pose estimates can drive reliable real-time feedback. In 2022, LearnOpenCV \cite{learnopencv_yolov7_mediapipe_2022} compared YOLOv7-Pose and MediaPipe in common pose estimation settings. In 2024, Roboflow \cite{roboflow_pose_models_2024} provided a deployment-oriented overview of widely used pose models, while Ultralytics \cite{ultralytics_yolo11_pose_2024} released documentation for a modern YOLO-Pose family emphasizing real-time performance. Additional 2024 comparisons and surveys examined strengths and limitations across OpenPose, MediaPipe, and YOLO-based variants \cite{saiwa_openpose_mediapipe_2024,sharma_comparative_pose_yolov7_mediapipe_2024,needham_pose_estimation_review_2024}. Importantly for rehabilitation, Zhang et al. \cite{zhang_mediapipe_yolov5_rom_2024} combined MediaPipe and a YOLO-based approach to support range-of-motion assessment, illustrating how pose estimation can be embedded into clinical measurement workflows.

Despite rapid progress, pose estimation can still degrade under clothing variation, suboptimal lighting, partial body visibility, or out-of-plane motion, which can bias joint-angle estimates and range-of-motion calculations. Habibie et al. \cite{habibie_in_the_wild_pose_cvpr_2019} emphasized the difficulty of ``in-the-wild'' pose estimation under occlusion and appearance variation. For rehabilitation systems, this motivates confidence-aware feedback strategies, temporal smoothing, and user guidance for camera placement and movement plane selection so that users are not misled by unstable joint estimates.

\subsection{Real-Time Feedback and Gamified VR Rehabilitation}
Rehabilitation systems that provide immediate, interpretable feedback can accelerate motor learning and support adherence \cite{youssef2024telehealth,gaafer2024immersive, attallah2024immersive}. Sigrist et al. \cite{sigrist2013augmented} reviewed augmented visual, auditory, and haptic feedback for motor learning and highlighted that feedback timing, modality, and clarity are central to skill acquisition. Exergaming literature has connected game mechanics to motivation and behavior change: Craig \cite{craig2013understanding} discussed understanding perception and action in sport using virtual reality technology, while Lohse et al. \cite{lohse2014virtual} reviewed VR therapy and the role of task design and feedback in rehabilitation outcomes. In a broader digital-health context, Sardi et al. \cite{sardi2017systematic} systematically reviewed gamification in e-health, while practical systems have explored biofeedback-driven avatars for guided rehabilitation \cite{ebert2015development} and more recent serious-game-based rehabilitation platforms \cite{saleh2025rehabilitative}.

Because VR rehabilitation is interactive, evaluation commonly combines functional outcomes with usability and experience measures. Peres et al. \cite{peres2020usability,peres2019usability_intechopen} demonstrated how usability results can be interpreted in rehabilitation-device contexts. For game-like VR applications, the Game Experience Questionnaire (GEQ) by IJsselsteijn et al. \cite{ijsselsteijn2013game}, and serious-games analyses for health contexts such as Wiemeyer and Kliem \cite{wiemeyer2012serious} help interpret engagement and motivation. Finally, access and deployability are recurring concerns for home rehabilitation; Ayed et al. \cite{ayed2019vision} reviewed exergames and VR systems for motor rehabilitation, and Souza-Lima and Rocha-Filho \cite{souzaliwa2022webrtc} discussed WebRTC-based approaches to cross-platform real-time interaction, which can reduce installation and compatibility barriers.

\paragraph{Research gap and contribution.}
Prior work establishes that VR can increase engagement in rehabilitation and that deep-learning-based pose estimation can quantify human movement. However, comparatively few end-to-end systems tightly integrate (i) low-cost, cross-platform VR delivery suitable for home use, (ii) robust real-time markerless pose tracking under home-environment constraints (occlusion, camera placement, lighting), and (iii) clinically interpretable feedback and progression metrics tailored to LBP exercise technique. Existing VR rehabilitation studies often emphasize feasibility and engagement without fine-grained, automated movement-quality assessment, while pose-estimation research frequently evaluates benchmarks without closing the loop to actionable rehabilitation feedback. Our work addresses this gap by coupling a VR-guided LBP exercise flow (grounded in common prevention and exercise guidance \cite{mayoclinic_backpain_prevention}) with real-time markerless tracking and feedback, enabling technique-aware guidance and longitudinal progress logging within a deployable system.

\section{System Overview}

\subsection{Design Goals and Constraints}
The primary objective of the TOSHFA system is to provide a clinically valid, accessible, and engaging alternative to traditional low-back pain (LBP) rehabilitation. To ensure therapeutic efficacy, the system is engineered to adhere strictly to physiotherapy principles, maintaining a rotation accuracy of $\pm2.1^{\circ}$ during spinal exercises. This clinical grounding is supported by a "home-safe" design philosophy, which utilizes real-time postural corrections to protect sedentary users from injury caused by improper form. By offering immediate feedback, the system allows patients to maintain fitness independence while performing routines in a domestic setting.

To bridge the gap in healthcare accessibility, the system is built upon a cost-effective hardware constraint, requiring only mid-tier consumer electronics such as a standard laptop with a 720p webcam and a smartphone paired with a budget-friendly ($<\$10$) cardboard-style VR headset. This low-cost approach is matched by a modular software architecture designed for cross-platform compatibility across Windows, Mac, and Linux. The system maintains a strict end-to-end latency constraint of less than 100ms to ensure the real-time responsiveness necessary for motor learning, while keeping CPU usage between 15–25\% to remain functional on standard hardware.

Furthermore, TOSHFA addresses the psychological barriers to rehabilitation—specifically the typical 40\% to 60\% adherence rate of conventional programs—by integrating gamification and behavioral psychology. Through the use of reward systems, including points, streaks, and achievement milestones, the system incentivizes consistent participation. To facilitate long-term clinical monitoring, all session metrics are logged in a human-readable JSON format, allowing therapists to review objective data on patient progress, rotation angles, and hold durations over time.

\subsection{Hardware Setup}
The TOSHFA prototype is engineered to utilize general-purpose, consumer-grade devices to maximize accessibility and ease of deployment. The primary computational and sensing component is a standard laptop or desktop computer, typically equipped with an Intel i5 processor and 8GB of RAM, which utilizes an integrated or external 720p camera to process user movements at 30 frames per second. For optimal pose estimation accuracy, it is critical that the camera is positioned between one and two meters from the participant to maintain a clear field of view.

To deliver an immersive rehabilitation experience without the need for specialized medical displays, the system employs a standard Android or iOS smartphone mounted within a low-cost, cardboard-style VR headset. This setup allows the smartphone to render stereoscopic images in a split-screen format, effectively transforming a personal mobile device into a virtual training environment. Connectivity between the processing unit and the display is maintained via a Wi-Fi-based UDP streaming protocol, which ensures synchronization by keeping latency below 10 ms over a standard home network, thereby eliminating the need for cumbersome physical cables.

The overall setup process is designed for high efficiency, requiring approximately five minutes to configure the webcam, establish wireless device connections, and initialize the Python and Unity software environments. By avoiding reliance on expensive sensors or high-end GPUs, the system provides a robust at-home rehabilitation platform that can be implemented in a standard domestic setting using existing hardware.

\subsection{Software Architecture}
The system's software architecture is built upon a modular client-server relationship, separating intensive data analysis from the user-facing display. A Python-based backend handles movement acquisition and logic, while a Unity frontend manages immersive 3D rendering and VR deployment. To ensure the real-time responsiveness essential for motor learning, these two components communicate via a low-latency User Datagram Protocol (UDP), which allows for high-speed data streaming without the overhead of traditional connection-based protocols.

\begin{figure}[t]
    \centering
    \includegraphics[width=\textwidth]{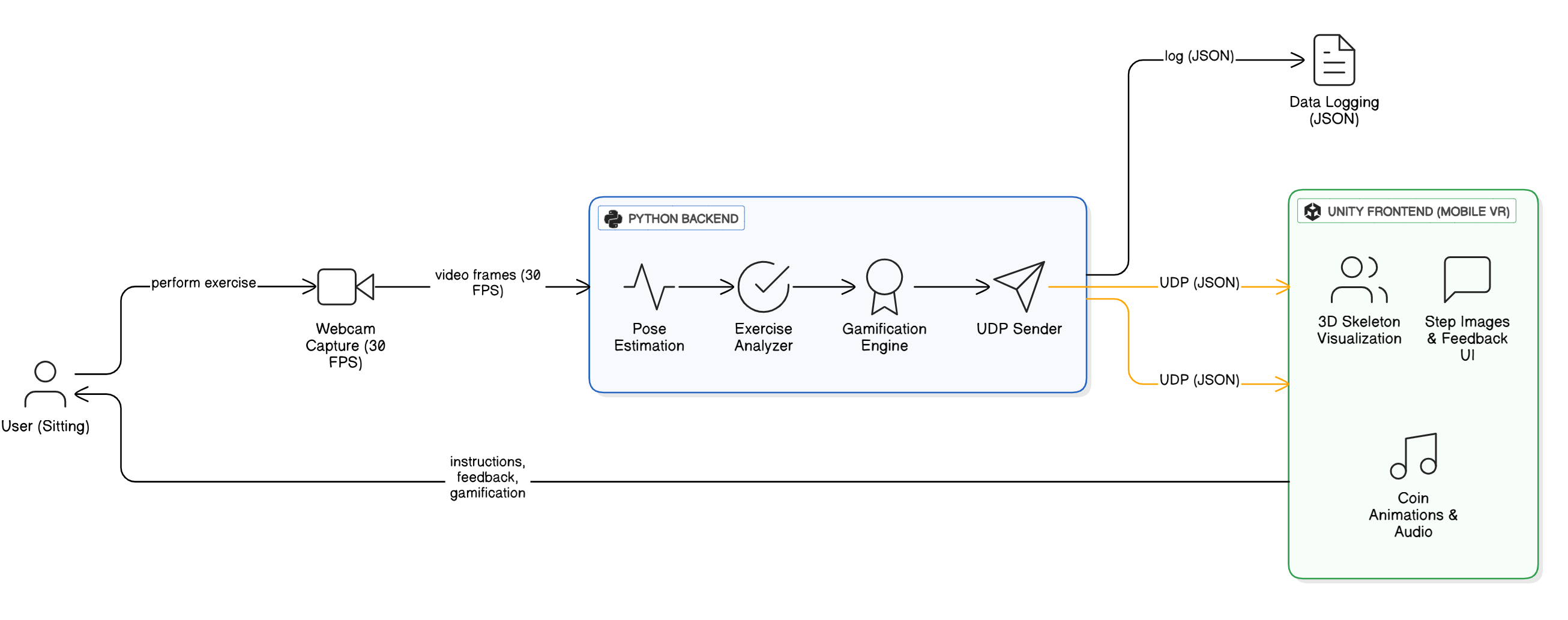}
    \caption{High-level overview of the TOSHFA software architecture, illustrating the low-latency UDP communication between the Python-based pose estimation backend and the Unity 3D mobile VR frontend.}
    \label{fig:software_architecture}
\end{figure}

\subsubsection{3.3.1 Pose Estimation Module}
For this system, the MediaPipe Pose algorithm was selected to perform real-time 3D landmark detection due to its superior performance on consumer-grade hardware. Unlike more resource-intensive alternatives like OpenPose, MediaPipe offers exceptionally low latency, typically under 20ms, while maintaining greater than 95\% accuracy across diverse body types. The module tracks 33 skeletal keypoints, with a specific focus on the shoulders (landmarks 11 and 12) and the hips (landmarks 23 and 24) to accurately calculate the torso rotation required for LBP exercises.

To ensure a stable user experience in home environments, the module processes 640x480 RGB frames at 30–35 FPS and implements temporal smoothing, such as exponential moving averages, to eliminate coordinate jitter. Furthermore, the system is robust against varying domestic lighting conditions, utilizing visibility scores with a threshold greater than 0.9 to filter out unreliable or false-positive detections. This results in an end-to-end inference pipeline that is highly responsive and optimized for CPU-based execution without requiring a dedicated GPU.

\subsubsection{3.3.2 Exercise Logic and Scoring Engine}
The scoring engine serves as the analytical bridge that transforms raw joint coordinates into actionable clinical insights. Torso rotation is determined by calculating the angles between the shoulder and hip vectors using the Arctan2 function, which achieves a measurement accuracy of $\pm2^{\circ}$. To ensure postural integrity, the system continuously verifies the user's "correctness" by checking the alignment of the shoulder and pelvis and monitoring the hip-to-knee ratio to confirm a proper sitting position. Exercises are deemed correct when rotation angles fall within a clinically safe range of $20^{\circ}$ to $60^{\circ}$; any movement exceeding this range is flagged to prevent potential injury

The engine operates through a state machine that progresses through neutral, right-rotation, and left-rotation phases. Feedback is calculated in real-time, assigning +10 points for correct repetitions and providing a +5 bonus for "excellence" when the user holds a rotation between $40^{\circ}$ and $50^{\circ}$. Conversely, a penalty of -5 points is applied if a postural fault is committed. The final output from this module includes a Boolean accuracy value, the specific rotation angle, a normalized hold-progress duration (0–1), and guided text prompts, such as "Rotate more to the right," to help the user adjust their form dynamically.

\subsubsection{Unity-Based Visualization and Mobile VR}

\begin{figure}[t]
\centering
\includegraphics[width=\textwidth]{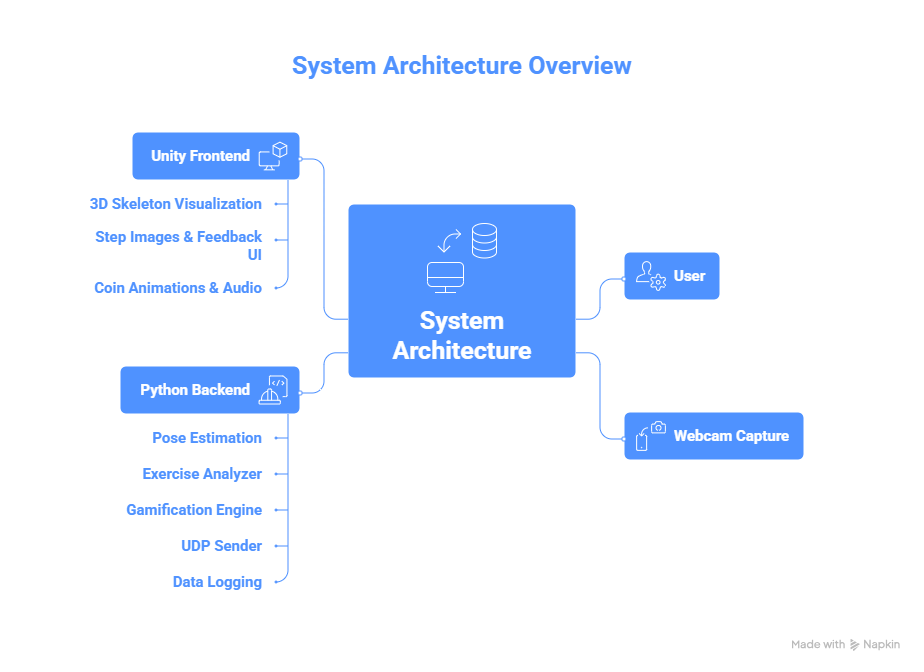}
\caption{Modular components of the Unity frontend architecture, detailing the integration of the 33-joint skeletal avatar, real-time feedback UI, and gamified reward mechanisms.}
\label{fig:unity_frontend}
\end{figure}

Unity 3D is utilized to render an immersive frontend specifically optimized for mobile VR deployment. The primary visual element of the Unity scene is a 3D avatar consisting of a 33-joint skeleton that mirrors the user's movements in real-time. To ensure a fluid and natural representation of motion, the system employs Lerp-interpolation for temporal smoothness. This avatar is supported by a comprehensive user interface that includes step-by-step exercise instructional images, dynamic progress bars, and real-time UI text providing feedback and procedural instructions.
Interaction within the virtual environment is guided by intuitive visual metaphors designed to reinforce correct form. For instance, joint materials dynamically switch between green and red to indicate postural correctness or faults. Furthermore, the system incorporates gamified rewards, such as coin animations paired with synchronized audio cues, to celebrate the successful completion of each repetition. This multimodal feedback loop is intended to enhance the sense of presence and encourage adherence through positive reinforcement. The overall architecture of this frontend is illustrated in the block diagram in Fig.~\ref{fig:unity_frontend}.

\subsubsection{Data Logging and Analytics}
The software architecture of the Seated Lower Back Rotational Stretch System adopts a modular, client-server design to ensure low-latency processing, real-time feedback, and scalability. It separates computation-intensive tasks, such as pose estimation and movement analysis, from the user-facing visualization layer, maintaining connectivity via a lightweight network protocol. The backend manages data acquisition and exercise logic, while the frontend is dedicated to immersive rendering. This architectural separation facilitates independent development, testing, and deployment, providing a robust foundation for potential cloud-based scaling in future iterations.

To support longitudinal clinical monitoring, the system records a comprehensive suite of logged metrics for each session. These metrics include start and end timestamps, detailed repetition-by-repetition data—specifically rotation angles, hold durations, and accuracy booleans—as well as aggregate performance indicators such as streaks, total scores, and specific events like posture errors. All data is organized within a structured JSON format to ensure maximum compatibility with external clinical databases and research tools. A representative example of this data structure is provided below:
\begin{verbatim}
{
"exercise": "seated",
"start_time": "2025-01-15T10:30:45Z",
"reps": [
{"rep_id": 1, "angle": 42.8, "hold_duration": 2.5, "correct": true}
],
"total_score": 50,
"streaks": 3
}
\end{verbatim}

\subsection{Exercise Design and Clinical Grounding}
\subsubsection{Exercise Selection}
The TOSHFA prototype is centered on non-invasive interventions designed to enhance lower lumbar spine (LBS) flexibility and alleviate chronic pain. The system's exercise protocols are grounded in evidence-based clinical practices, drawing from peer-reviewed research and the official guidelines of the American Physical Therapy Association (APTA). By integrating over 15 scientific references supporting non-invasive lumbar care, the system ensures that the digital rehabilitation experience aligns with traditional physical therapy standards.

For this pilot study, the seated lower back rotational stretch was selected as the primary therapeutic movement. This exercise is performed with the user seated upright in a chair, feet flat on the floor, and the spine in a neutral position relative to the pelvis. The protocol requires a controlled rotation between $20^{\circ}$ and $60^{\circ}$ on alternating sides, with each stretch held for a duration of 2–3 seconds. To ensure patient safety and therapeutic compliance, the system utilizes a hip-to-knee ratio to verify proper posture and prevent "cheating" or standing during the routine. Crucially, the system monitors the range of motion to ensure users do not exceed $60^{\circ}$ of rotation, as over-rotation significantly increases the risk of lumbar injury. By requiring no specialized equipment, this guided exercise serves as an accessible solution for reducing stiffness in sedentary office workers, the elderly, and post-surgical patients.

\subsubsection{Gamification and Feedback Mapping}
The TOSHFA system ensures that exercise correctness has a direct and immediate impact on motivational game elements, leveraging behavioral psychology to foster adherence. Central to this experience is a dynamic progress bar that fills during isometric holds, using color-coded feedback—green for accurate positioning and red for incorrect form—to guide the user through the 0–1 hold duration. Upon the successful completion of a repetition, the system triggers collectibles in the form of coin animations accompanied by rewarding audio cues, granting the user +10 points. To further incentivize precision, performance bonuses are awarded when the user maintains rotation within optimal clinical angles.

Long-term engagement is sustained through a structured system of streaks and achievements designed to celebrate session milestones. For example, users can unlock badges such as the "5-Minute Warrior" for sustained activity, while maintaining a series of consecutive correct movements builds a streak; reaching a five-repetition threshold triggers a "Perfect Streak" popup accompanied by celebratory fanfare.

This motivational framework is reinforced by an encouragement system that maps exercise states to specific text and audio prompts. Achieving a "Perfect RIGHT rotation!" is met with a positive sound and affirming text, whereas postural errors trigger a guiding tone and corrective instructions, such as "Rotate more to the right →". By carefully balancing these rewards and corrections, the system utilizes operant conditioning to build long-term adherence without over-rewarding the user.

\section{Methods}

\subsection{Study Design}
The Pilot Study was a small sample study with one session to test the usability and user experience of the AI-based physiotherapy. The Pilot Study was conducted in a lab environment in which the participants performed a number of physiotherapy exercises while receiving real-time feedback regarding their poses and movements via video analysis.

The participants were seated comfortably in the lab environment during the testing session. The participants were monitored using a mobile phone video camera, which streamed the video feed to a laptop that processed the video feed and showed the participants an Eleven-Eleven application. This application displayed exercise performance instructions and visual feedback during the performance of the exercises. The application runs on a laptop computer, and it allows the participants to monitor their progress with exercises and receive real-time feedback based on how well they were performing the exercises.

\begin{figure}[t]
    \centering
    \includegraphics[width=\textwidth]{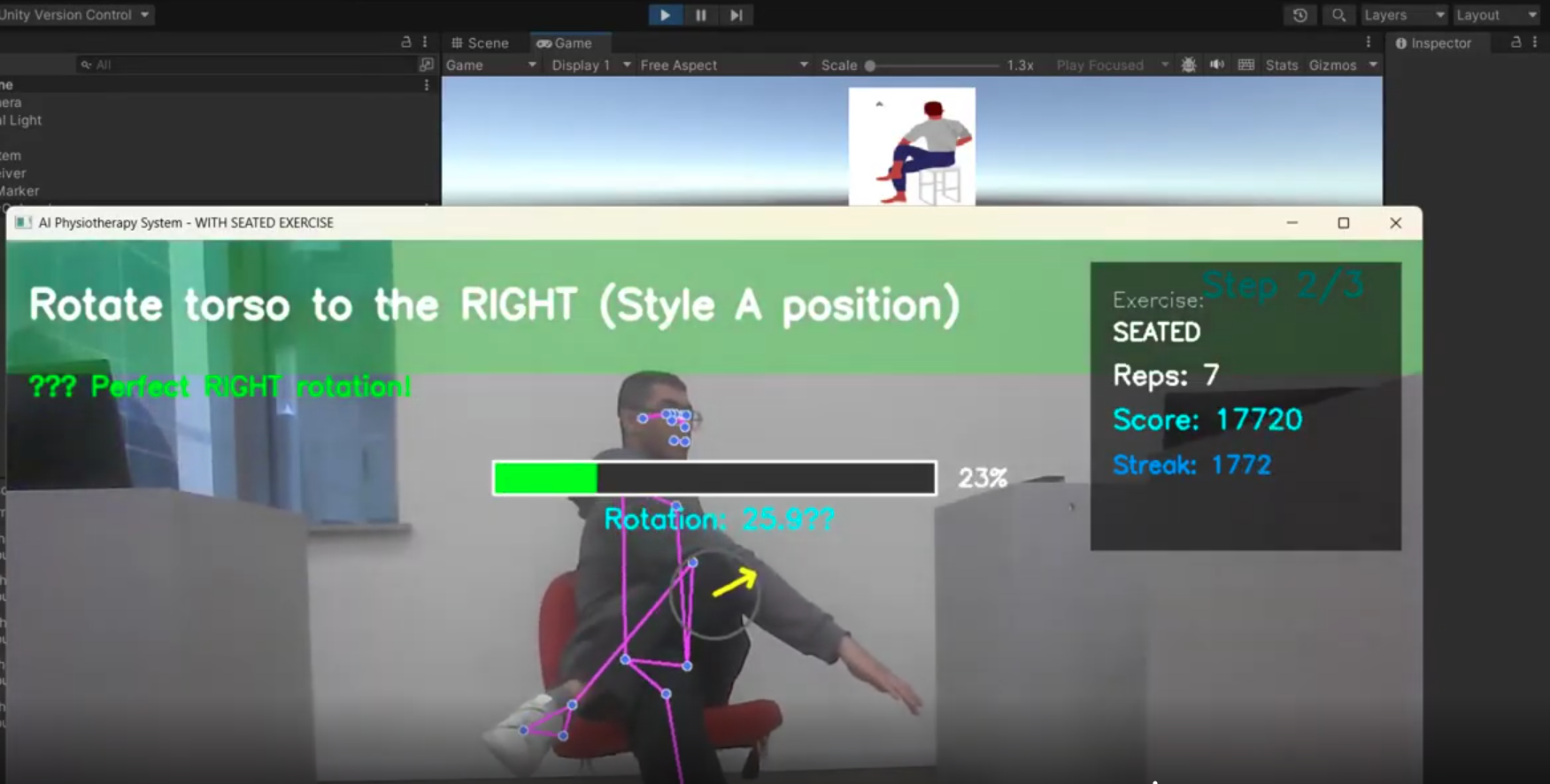}
    \caption{Experimental setup and research team environment during the pilot usability study sessions.}
    \label{fig:team}
\end{figure}

\subsection{Participant Selection and Demographics }

The study established specific inclusion criteria to ensure a baseline level of physical capability and technological literacy among the cohort. Eligible participants were required to be between 18 and 23 years of age. Clinically, candidates had to be free of any major musculoskeletal or neurological conditions that might impede basic rotational motions. Furthermore, participants were required to demonstrate the cognitive ability to comprehend and follow simple verbal instructions, alongside a functional familiarity with smartphone operation.

Conversely, individuals were excluded from the study if they presented with moderate to severe low back pain or any other physical condition that precluded safe exercise. Additional exclusion criteria included pregnancy, a history of joint replacement surgery, or recent fractures of the hand or ankle. Finally, because the exercise protocol requires sustained torso rotation, individuals unable to tolerate sitting for a prolonged period were also excluded from participation.

A total of twenty individuals ($N=20$) were recruited for the usability and experience trials. The cohort was characterized by a balanced gender distribution, consisting of 10 males and 10 females, with a mean age of 20 years and a range of 18 to 22 years. The participants were primarily healthy with no significant underlying medical issues. However, consistent with the study's focus on LBP rehabilitation, some individuals within the sample reported mild to moderate low back pain as their primary medical condition.

\subsection{Procedures}

\subsubsection{Participant Onboarding}
The onboarding process was designed to ensure that all subjects were fully informed and physically prepared before commencing the trials. Initially, participants were presented with an informed consent document that explicitly detailed the study's primary objectives, potential physical risks, and the protocols for data privacy and usage. Following the formal consent, a comprehensive safety briefing was conducted to ensure that participants understood the nature of the required physical activities and were instructed on how to report any discomfort or distress during the session.

The final stage of onboarding focused on system familiarization, where participants received a guided introduction to the Unity application. This included an overview of how the app delivers exercise instructions and tracks individual progress. To ensure operational competency, researchers provided a demonstration of the system's interaction mechanics, specifically showing participants how to manipulate the interface and respond to real-time feedback through postural adjustments and timed holds.

\subsubsection{Exercise Protocol, Researcher Monitoring and Intervention, and User Feedback Interface}
The exercise session consisted of a structured sequence of trials focusing on three core movements: the seated rotation (a chair-based exercise), the bridge, and the cat-cow. While the specific number of repetitions per set was not predetermined, each individual repetition was performed for a duration of 15 seconds without intervening rest periods. The total duration of a session typically ranged between 5 and 8 minutes, varying based on the participant's individual pace and the elective resting time taken between separate trials.

Throughout the session, participants engaged with the system through a smartphone screen positioned directly in front of them to receive continuous real-time feedback. The visual interface provided explicit movement prompts, such as "rotate to the left," alongside a dynamic progress bar that tracked the completion status of each exercise. To encourage adherence, the system displayed verbal praise, including messages like "Good position; rotate more". While full audio-guided prompts were not utilized in this protocol, optional auditory cues, such as a completion beep, were integrated to enhance interactivity and signal the end of successful repetitions.

The researcher monitored participants from a computer screen showing Unity's interface. If it became obvious that a participant was having difficulty executing an exercise, the researcher provided verbal assistance to correct or improve their execution. The researcher was able to monitor real-time pose information to allow interaction with participants (e.g., "rotate too far left; try to maintain a central position").

\begin{figure}[t]
    \centering
    \includegraphics[width=\textwidth]{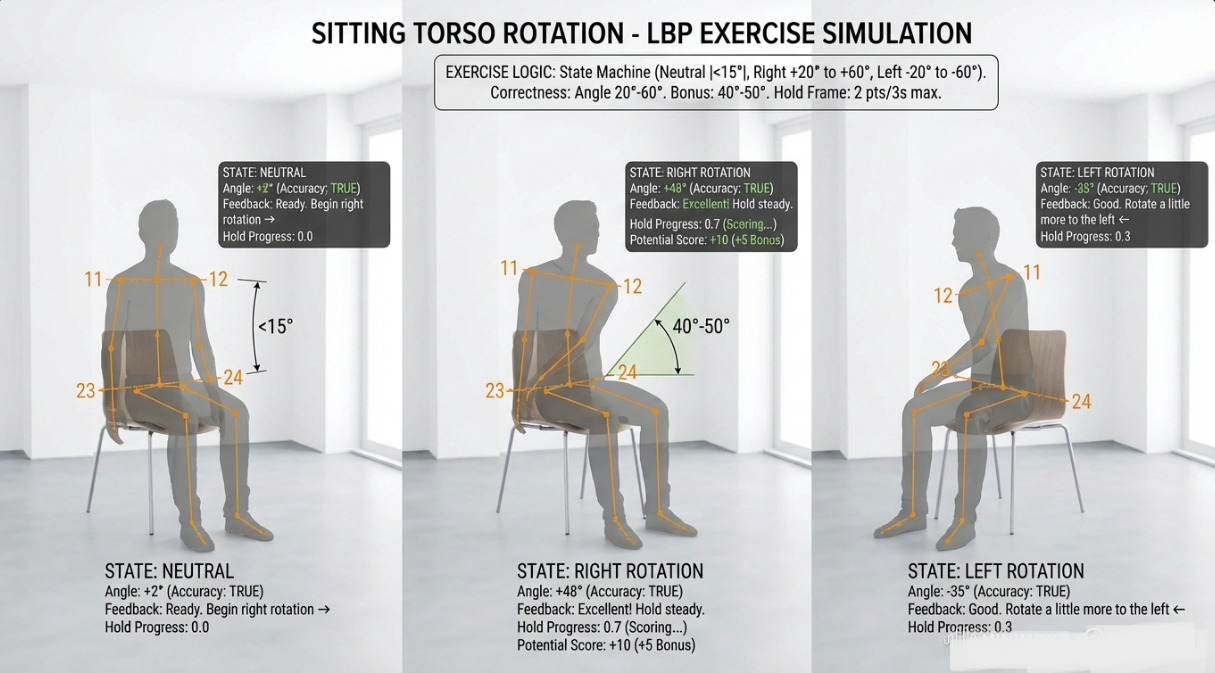}
    \caption{Visualization of the "Seated Torso Rotation" state machine, demonstrating the system's ability to provide real-time postural correction and angle-specific scoring bonuses.}
    \label{fig:perfectn}
\end{figure}

\subsubsection{Post-Session Questionnaires and Interviews}
Upon completion of the exercise session, participants underwent a multi-dimensional assessment to quantify both system usability and the psychological quality of the user experience. The primary metric for functional evaluation was the System Usability Scale (SUS), a standardized 10-item instrument used to calculate a usability score ranging from 0 to 100, where higher values represent superior system ease-of-use.

To complement the usability data, participants completed a customized Game Experience Questionnaire (GEQ) designed to capture the hedonic and motivational aspects of the physiotherapy platform. This assessment examined six key experiential dimensions through targeted statements: Competence ("I feel that I am getting better at this exercise"), Immersion ("I felt that I was fully engaged in the exercise session"), Flow ("The exercise session was seamless and uninterrupted"), Positive Affect ("I liked the feedback and interactions with the system"), Negative Affect ("There were some parts of the system that I found frustrating to use"), and Tension ("I felt overwhelmed or stressed by the instructions").

Finally, participants were invited to participate in an optional semi-structured interview to provide open-ended qualitative feedback. This dialogue allowed researchers to explore specific user preferences and pain points through three guiding questions: 1) what was your favorite thing about the exercise system?, 2) what was your least favorite thing about the exercise system?, 3) and were there moments throughout the exercise session when you felt confused or frustrated?

\begin{figure}[t]
    \centering
    \includegraphics[width=\textwidth]{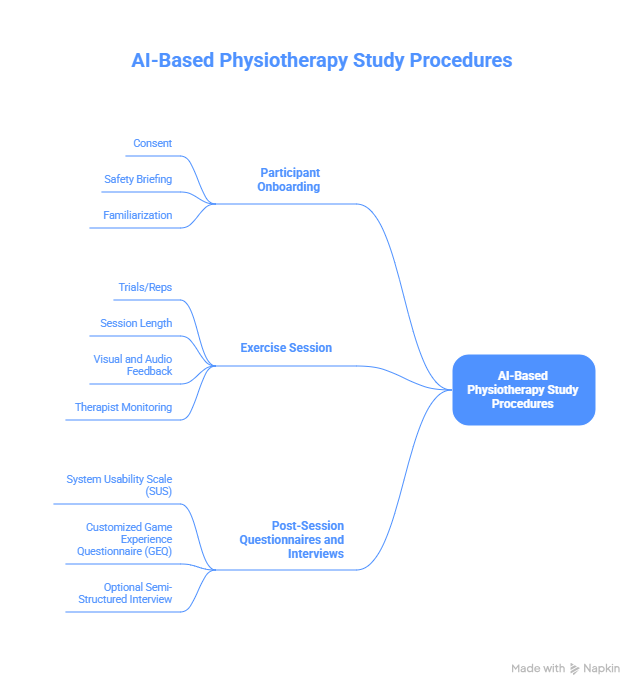}
    \caption{Methodological workflow of the pilot study, depicting the sequential stages of participant onboarding, supervised exercise trials, and multi-metric post-session evaluations.}
    \label{fig:procedures}
\end{figure}

\subsection{Quantitative and Qualitative Data Analysis}
The data analysis for this study utilized a mixed-methods approach to provide a comprehensive evaluation of the TOSHFA system. Quantitative assessment began with the calculation of System Usability Scale (SUS) scores, where individual results were derived for each participant and subsequently averaged to determine the overall mean SUS score. For the Game Experience Questionnaire (GEQ), response data were aggregated into specific dimensions—including Competence, Immersion, and Flow—to calculate both the mean dimension scores and the overall mean GEQ score. Additionally, standard deviations were calculated for these metrics to evaluate the variability of participant responses and the consistency of the user experience.

Beyond subjective reporting, the study analyzed objective performance metrics by computing the accuracy of each participant's joint angles relative to the clinically ideal positions defined in the exercise logic. These data were reported as a percentage accuracy value, averaged across participants for each specific exercise. To complement these quantitative findings, qualitative data derived from the semi-structured interviews underwent an open coding process. This text was thematically grouped based on participant feedback regarding user experience, usability, and engagement. This thematic analysis allowed for the identification of specific participant preferences and critical areas for system improvement, providing a roadmap for future iterative design.

\begin{figure}[t]
    \centering
    \includegraphics[width=\textwidth]{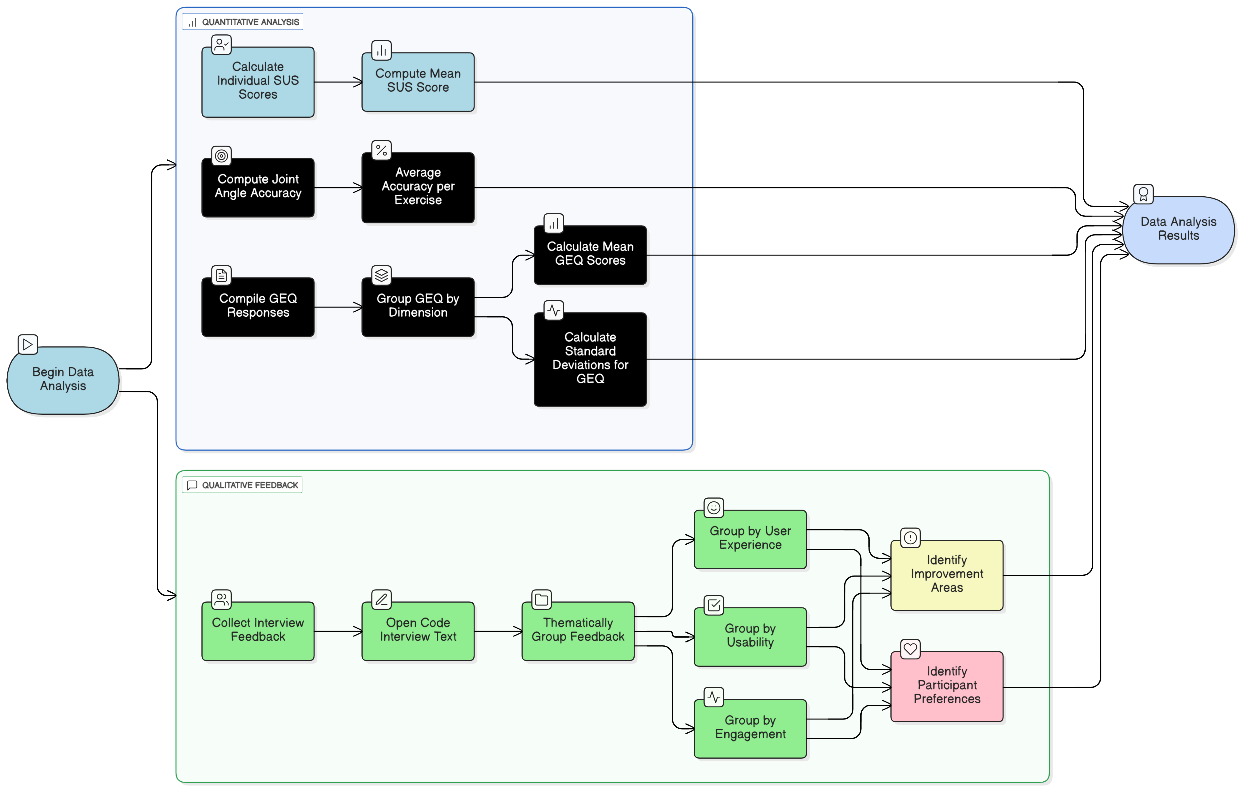}
    \caption{Mixed-methods data analysis pipeline, illustrating the quantitative processing of SUS/GEQ scores and the qualitative thematic coding of participant interviews.}
    \label{fig:img}
\end{figure}


\section{Results and Discussion}

\subsection{Interpretation of Usability and Experience Outcomes}
The pilot evaluation indicates that while the prototype maintains functional feasibility, its usability currently sits below the industry standard. Analysis of the SUS results yielded an overall mean score of 47.4 ($SD \approx 7.1$). This value remains significantly below the established benchmark of 68, which serves as the threshold for "good" usability, placing the current system within the "poor-to-marginal" classification. The variability observed between individual scores—ranging from a minimum of 27.5 to a maximum of 60.0—suggests that participants encountered recurring obstacles when attempting to complete exercise tasks. This technical friction likely contributed to the frustration reported by the younger cohort (ages 19–22), who typically demand high levels of intuitive design and seamless functionality.

However, these findings must be contextualized within the early-stage prototype condition of the TOSHFA system. It is common for initial ratings of VR-based rehabilitation tools to be lower than final versions, as users require time to adapt to novel interaction paradigms and specialized clinical interfaces. Despite the lower scores, the low variance in SUS ratings suggests a consistent experience across the cohort, indicating that the challenges were systemic rather than isolated failures. Furthermore, the gender-balanced sample (50

In contrast to the usability challenges, the GEQ results revealed high levels of experiential engagement. Participants reported strong positive emotional responses, with high mean scores in dimensions such as interestingness and enjoyment (e.g., $Q1=4.7$, $Q9=4.5$, $Q13=4.95$, and $Q14=5.0$). These data suggest that the system succeeded in being emotionally engaging despite the presence of "interaction friction". Lower scores in specific subcomponents, such as flow and immersion (e.g., $Q12=2.7, Q4=3.0$), pinpoint areas where system limitations or setup requirements may have momentarily interrupted the user's sense of presence.

\subsection{Design Implications for Mobile VR Rehabilitation Systems}
The results of this pilot study provide several critical design implications for the evolution of mobile VR rehabilitation platforms. A primary observation is the significant disparity between the marginal SUS usability scores and the high GEQ engagement ratings. This suggests that while technical friction exists, the "fun factor" of gamification maintains user interest. To bridge this gap, future iterations must prioritize the reduction of interaction complexity by simplifying navigation menus and minimizing the number of discrete actions required to initialize a workout. Streamlined onboarding processes and clearer instructional sets are likely to enhance usability without compromising the hedonic qualities that drive patient motivation.

Furthermore, the high ratings for enjoyment and positive affect—despite the system's current limitations—indicate that early-stage rehabilitation can be effectively delivered through relatively simple feedback mechanisms, such as basic sound effects and visual avatars. Designers should be cautious of over-complicating the interface with excessive sensory feedback, as high visual or auditory complexity may increase the user's cognitive load. This is particularly critical for patients suffering from chronic pain, whose cognitive resources may already be taxed by the physical demands of the exercise.

The moderate immersion scores also highlight the technical necessity of tracking stability and precise camera calibration. In a mobile VR context, even minor discrepancies in pose estimation or tracking latency can disrupt the user's "flow," making it difficult to maintain the connection between their physical movements and the virtual environment. Consequently, mobile VR developers should adopt a conservative design approach that emphasizes tracking robustness and interface usability over high-fidelity visual complexity.

Finally, safety remains the paramount concern for independent, home-based rehabilitation. The architecture must strike a delicate balance between providing an immersive experience and ensuring the user remains aware of their physical surroundings. Future designs may benefit from "safety-first" features, such as proximity alerts or semi-transparent UI elements, to prevent injuries when patients are exercising without professional supervision.

\subsection{Limitations and Future Directions}
While this pilot study establishes the initial feasibility of the TOSHFA system, several limitations must be addressed in future iterations to ensure clinical and statistical generalizability. The current sample size is relatively small, and the cohort consists entirely of young adults (ages 19–22) with a high degree of technological literacy. Consequently, these findings may not fully represent the primary target populations for LBP rehabilitation, such as elderly individuals or post-surgical patients, whose physical capabilities and familiarity with VR may differ significantly. Furthermore, because assessments were conducted over a single session in a controlled laboratory environment, the study could not evaluate long-term adherence, learning effects, or the system's performance in authentic home-usage scenarios.

To address these constraints, future research will focus on expanding both the participant demographic and the clinical scope of the system. We plan to conduct usability refinements—including interface simplification and enhanced onboarding—based on the feedback gathered in this trial. Subsequent studies will transition from healthy volunteers to clinical populations, assessing elderly and recently injured participants across multiple rehabilitation exercises. To evaluate therapeutic effectiveness, future protocols will integrate longitudinal monitoring and objective clinical outcome measures, such as pain scores, functional disability indices, and patient-reported outcomes.

Ultimately, the results of this pilot define the technical foundation and identify the major design challenges required to move toward a larger, more comprehensive clinical trial. By evaluating home deployment and remote monitoring, we aim to validate TOSHFA as a robust, real-world solution for accessible low back pain management.

\section{Conclusion}
Low back pain (LBP) remains a pervasive global health challenge, impacting a vast majority of the adult population and incurring high socioeconomic costs. Traditional rehabilitation frameworks are frequently hindered by systemic barriers, including limited clinical accessibility, poor patient compliance, and a lack of continuous professional monitoring. To address these issues, this work introduced a novel, mobile Virtual Reality (VR) rehabilitation system that leverages webcam-based pose estimation to provide immediate biofeedback. By utilizing consumer-grade hardware, the TOSHFA system achieves a level of affordability and scalability that allows for effective, home-based LBP management.

The primary contributions of this research include the development of a clinically grounded, pose-guided architecture and the completion of a pilot user experience study with a cohort of 20 participants. While results indicated moderate usability levels—highlighting a clear need for further interface refinement—participants reported high levels of enjoyment and motivation during the sessions. These findings suggest that the integration of gamified elements successfully offsets technical friction, providing a robust foundation for enhancing future user interaction.

Ultimately, this study demonstrates the feasibility of combining markerless pose estimation with mobile VR to create an accessible telerehabilitation paradigm. This work lays the critical groundwork for future research, which will focus on optimizing system usability, expanding the library of integrated exercises, and launching longitudinal clinical trials to evaluate long-term therapeutic outcomes.

\bibliographystyle{IEEEtran} 
\bibliography{references}

\end{document}